# The Evolution of Reputation-Based Cooperation in Regular Networks

Tatsuya Sasaki [1,*], Hitoshi Yamamoto [2], Isamu Okada [3] and Satoshi Uchida [4]

[1] Faculty of Mathematics, University of Vienna, 1090 Vienna, Austria
[2] Faculty of Business Administration, Rissho University, 141-8602 Tokyo, Japan
[3] Faculty of Business Administration, Soka University, 192-8577 Tokyo, Japan
[4] Research Center for Ethiculture Studies, RINRI Institute, 102-8561 Tokyo, Japan
* Correspondence: tatsuya.sasaki@univie.ac.at; Tel.: +43-1-4277-50774

**Abstract:** Despite recent advances in reputation technologies, it is not clear how reputation systems can affect human cooperation in social networks. Although it is known that two of the major mechanisms in the evolution of cooperation are spatial selection and reputation-based reciprocity, theoretical study of the interplay between both mechanisms remains almost uncharted. Here, we present a new individual-based model for the evolution of reciprocal cooperation between reputation and networks. We comparatively analyze four of the leading moral assessment rules—shunning, image scoring, stern judging, and simple standing—and base the model on the giving game in regular networks for Cooperators, Defectors, and Discriminators. Discriminators rely on a proper moral assessment rule. By using individual-based models, we show that the four assessment rules are differently characterized in terms of how cooperation evolves, depending on the benefit-to-cost ratio, the network-node degree, and the observation and error conditions. Our findings show that the most tolerant rule—simple standing—is the most robust among the four assessment rules in promoting cooperation in regular networks.

**Keywords:** evolution of cooperation; indirect reciprocity; structured population; social norm; private information

**PACS Classifications:** 87.23.-n; 02.50.Ey

**JEL Classifications:** C72; C73; D63; D64; D82; D85

## 1. Introduction

Reputation is one of the most practical tools for measuring partners' quality and incentivizing partners' behaviors [1]. Reputation is thus often compared to currency [2]. The concept of a reputation system has been applied to various situations, e.g., from gossip among neighbors to a rating and review for e-Bay (San Jose, CA, USA), Uber (San Francisco, CA, USA), TripAdvisor (Needham, MA, USA), etc. Game-theoretical studies have shown that reputation can facilitate the evolution of reciprocal cooperation in the context of indirect reciprocity [3–6]. Indirect reciprocity through reputation works in a peer-to-peer fashion by considering conditional cooperation: to help his/her co-player who has a good reputation yet also refuse to help the co-player who has a bad reputation [7]. The crucial aspect of reputation-based indirect reciprocity is how individual profiles are assessed in terms of their image score or morally judged as being good or bad [8,9]. The mapping of individual profiles to the image score is called the moral assessment rule [10]. Both classification and analysis of the moral assessment rule in the situation of social exchange have attracted broad attention in the fields of evolutionary biology and the social sciences [7].

In this study, we shed light on the effects of population structures on the evolution of reputation-based indirect reciprocity. Spatial selection is another major factor of the evolution of cooperation [11–14]. There is a vast amount of game-theoretical literature on the evolution of cooperation through direct reciprocity [15–24] or upstream reciprocity [25–27] in structured populations. Here, we consider networks of residents in which the reputations of neighbors are key pieces of information used to determine not only the move but also the partner in the next interaction. This allows us to explore an indirect interaction, which can be described as follows: "I know you did not help a good resident in the past (and thus you look bad); therefore, today I will not help you". Although this situation is quite popular in real-life settings, it has had little exploration in game-theoretic models. In this paper, we consider that the focal player's action is determined not only by how the neighbor behaved to the focal player but how the neighbor behaved to the neighbor's neighbor. Similarly, the neighbor's last action to the neighbor's neighbor may be determined by how the neighbor's neighbor previously behaved to the neighbor's neighbor's neighbor. Thus, reputation is good at both accumulating and abstracting this information, and population structures may affect not only the shape of individual interaction within a neighborhood but also the formation of individual reputational information.

Positive effects of the interplay of population structures and reputation-based conditional behaviors as such have been numerically investigated since approximately 2000 [28–30]. Models in previous studies mostly combined punishment, partner choice, or network rewiring [31–37]. Notably, punishment, partner choice, and network rewiring are often costly [38,39].

In this study, we would like to break into an unexplored area of the moral assessment rules. To date, the assessment rules explored for spatial indirect reciprocity have included only the simplest—the image-scoring rule [40]. Image scoring depends only on the focal player's last action, which thus is called the first-order assessment rule [10]. Recent experimental evidence has shown that a certain fraction of people is likely to use not only the information of the focal individual but also the opponent's profile [41]. Assessment rules that consider the opponent's reputation as well as the focal player's last action are called second-order assessment rules [10]. Second-order assessment rules may require more cognitive loads and have higher information costs than image scoring [42,43]; thus, they may be more likely to invite those who freeload on others' efforts in assessment [44]. However, recent advances in information and communication technology (ICT) are lowering the threshold of applying such complicated assessment rules in indirect reciprocity. An institutional pre-assessment system can also help deter the assessment freeloader [45].

The major second-order assessment rules consist of simple standing [3,46,47], stern judging [48,49], and shunning [50] (Table 1). However, little is known about how these second-order rules affect the evolution of cooperation by spatial indirect reciprocity [51]. Therefore, the primary aim of this paper is to comparatively analyze the four representative assessment rules of simple standing, stern judging, shunning, and image scoring in evolutionary games on social networks. We base evolutionary giving games on regular networks and consider public and private information with assessment errors. As we will show in the following chapters, our model can lead to clearly distinguishing among the four representative assessment rules. Our results reveal that simple standing, which is the most tolerant among the four rules, is the most robust in sustaining full cooperation; in the other three rules, cooperation becomes less frequent and commonly does so as the node degree increases and the assessment error becomes private.

In the following sections, we propose an agent-based model for studying spatial indirect reciprocity (Section 2), numerically analyze the evolution of spatial indirect reciprocity with second-order assessment rules (Section 3), and discuss possible reasons, applications, and implications of our results (Section 4).

**Table 1.** What is good and what is bad? "G" and "B", respectively, describe a good and bad image, and "C" and "D", respectively, describe offering to help and refusing to help.

| Conditions | Image of recipient | G | G | B | B |
|---|---|---|---|---|---|
|  | Action of donor | C | D | C | D |
| Assessment rule: What does the donor's image look like? | Shunning (SH) | G | B | B | B |
|  | Stern judging (SJ) | G | B | B | G |
|  | Image scoring (IS) | G | B | G | B |
|  | Simple standing (ST) | G | B | G | G |

## 2. Materials and Methods

We will first consider evolutionary giving games in finite structured populations. As in the "spatial indirect reciprocation" model [28], each individual plays giving games only within a given neighborhood and updates his/her own strategy through the pairwise payoff comparison with a random neighbor. In this study, we examine second-order assessment rules, as mentioned earlier. Conditional behaviors of each individual can thus be influenced by behaviors of remote third parties that are not in the neighborhood.

*2.1. Individual-Based Model*

Population Structure, Individual Structure, and Trial Sequence

**Regular ring lattice.** We consider $N = 400$ (fixed) individuals. We assume that all individuals are placed randomly on nodes of the regular ring lattice (Figure 1). The number of nodes equals that of individuals, and there is only one individual per node. The node degree of the regular ring lattice is given by an even number $k$ so that each node connects to all its nearest neighboring nodes, and the number of the connections per node is $k$. The connections of nodes are fixed, and the locations of individuals are unchanged throughout a simulation trial.

**Individual structure.** Each individual has the following basic attributes: {id, location, payoff, strategy, self-image, others image list}. Each individual adopts a specific strategy among the three strategies {Cooperator (ALLC), Defector (ALLD), Discriminator (DISC)}. Each individual can be assigned different image scores by others because they may have different assessment rules or make errors in private assessment. Every individual will update his/her image-list of all other individuals in the population [52,53]. We particularly assume that self-image is fixed as "good" and unchanged throughout a trial of individual-based simulation.

**Simulation trial.** One trial of the simulation consists of $g = 500$ generations, and each generation consists of $h = 50$ periods. In the simulation trial, the individual strategy is initially given at random from the three available: ALLC, ALLD, and DISC. We assume that, at the beginning of every generation, the image scores of all individuals are initialized as good. In each period, every individual as a donor commits once to a giving game in random order. Thus, $N = 400$ game-player turns occur per period. At the end of every generation, all individuals synchronously update their own strategy.

Each game-player turn comprises the following three phases:

(1) **Player-selection phase.** A focal player is selected, as aforementioned, and is then offered an opportunity to help a recipient player, who is randomly selected from the focal player's closest neighborhood.
(2) **Giving-game phase.** This is a one-shot giving game [7]. Depending on his/her strategy (whose details are given later), the focal individual determines whether to give help to the recipient or not. Giving help requires either personal cost $c > 0$ or nothing. Giving help means to play C, and not-giving help means to play D. Each helping action leads to benefits $b$ for the recipient with $b > c$. This is a social-dilemma situation: if the interaction is random matching, irrespective of what others do, switching to playing D is more advantageous than playing C by saving costs $c$; nevertheless, the net payoff is 0 if both play D and $b - c > 0$ if both play C. We assume *implementation errors*, in which the focal player who intends to play C will implement D with probability $p$ and, similarly, the focal player who intends to play D will implement C with probability $p$. That is, the implementation error is bilateral.
(3) **Image updating phase.** Finally, each player (except for the focal player) synchronously updates his/her own player-image list by assessing the focal player. We examine two extreme monitoring scenarios: every giving game is monitored by (i) a representative observer with a proper assessment rule (indirect observation) or (ii) all players (except for the focal player) (direct observation) [54]. In (i) indirect observation, the representative observer assesses the focal player, relying on the focal player's last action in the giving game and the recipient's image. We assume *assessment errors*: in making assessments, the representative observer makes errors with probability $q$, in which the representative observer assigns a good image to those who, in the case with no assessment error, should have a bad one or a bad image to those who, in the case with no assessment error, should have a good one. Hence, the assessment error is bilateral. The same assessment information regarding the focal player is then shared by all individuals, whether that information is erroneous or not [8,9]. In (ii) direct observation, all observing individuals independently assess the focal player, and each individual independently commits to assessment errors with probability $q$ [55], as is assumed of the representative individual in (i). In this study, we do not consider any specific consensus formation among individuals.

**Strategy updating and mutation.** We assume that all individuals undergo probabilistic strategy updating and rare mutation synchronously at the end of every generation. For strategy updating, the model individual is randomly chosen among the closest neighbors of the focal individual. As with replicator dynamics in well-mixed populations [56], we consider the selection, which depends on the payoff difference between the two. Let $P_i$ be the accumulated payoff of individual $i$ throughout the last generation. The probability for a focal individual $i$ to select the model $j$'s strategy is defined as:

$$\Pr(i \to j) = 1/[1 + \exp\left(-s(P_j - P_i)\right)] \tag{1}$$

The focal individual then undergoes mutation with probability $m$ and, if so, the focal individual will randomly switch to one of the given three strategies.

*2.2. Game Strategies and Assessment Rules*

To investigate the effects of different assessment rules on the emergence of indirect reciprocity in social networks, we consider the following typical strategies:

- Defector (ALLD): playing D unconditionally
- Cooperator (ALLC): playing C unconditionally
- Discriminator (DISC): playing C (if the recipient has a good image) or playing D (if the recipient has a good image)

We assume that, in the beginning of individual-based simulation, each node of the regular ring lattice is occupied with a strategist that is randomly selected from the above three, unless otherwise instructed.

Who is good or bad is determined by the assessment rule. In this paper, we examine the four assessment rules, as follows (see also Table 1):

- Shunning (SH): either assessing a donor as good if the donor plays C to a recipient who has a good image or assessing the donor as bad. Shunning is the strictest among the four rules [50].
- Image scoring (IS): either assessing a donor as good if the donor plays C or assessing a donor as bad if the donor plays D to a recipient, irrespective of the recipient's image. Image scoring is the simplest among the four rules because it depends only on the donor's action [40].
- Stern judging (SJ): assessing a donor as good if the donor either plays C to a recipient who has a good image or plays D to a recipient who has a bad image. Stern judging is the second strictest assessment rule because a player who has a bad image can also cleanse that image by refusing to help another player who has a bad image [48,49]. This assessment of defection is a so-called "justified defection" [3]. Stern judging is one of the eight leading rules [8,9].
- Simple standing (SS): either assessing a donor as bad if the donor plays D to a recipient who has a bad image or assessing a donor as good [3,46,47]. Simple standing is the most tolerant among the four rules and is also one of the eight leading rules.

Figure 1 shows an example of a series of game interactions and image assessments.

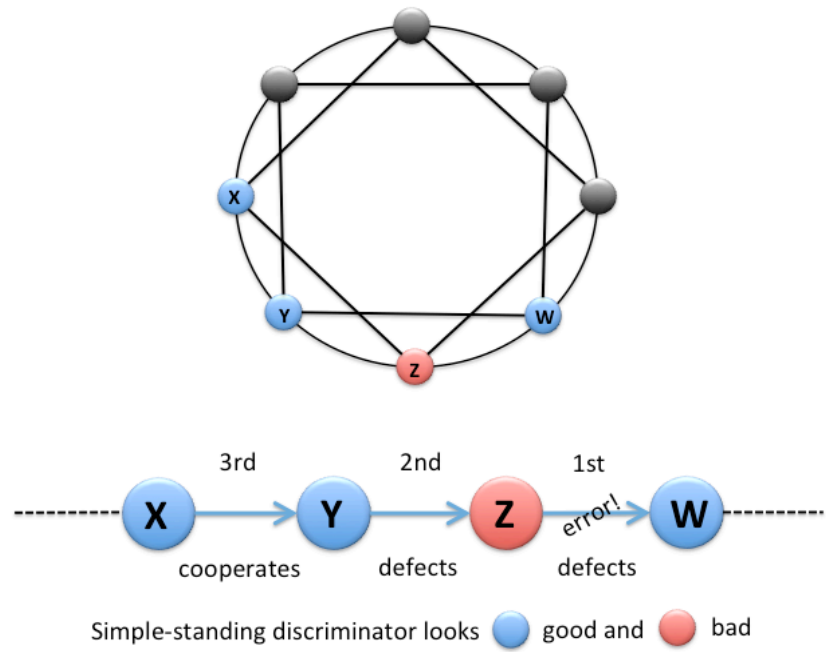

**Figure 1.** Actions and assessments in the giving games in a regular network with a population size of 8 and a node degree of 4. We assume that individuals X, Y, Z, and W in the network above are all Discriminators; X, Y, and Z adopt the simple standing rule and, initially, W has a good image under simple standing. The order of play is as follows: first, Z (donor) to W (recipient), second, Y to Z, and third, X to Y. It will follow that, first, Z intends to cooperate, and we assume that Z erroneously defects to W. Thus, the image of Z becomes bad from the viewpoint of ST. Second, Y will defect to Z; thus, X's image of Y will become good. Finally, X will cooperate with Y.

## 3. Results

In Figures 2 and 3, we numerically calculate the rate of the cooperation action C over all actions as controlling benefits $b$, node degrees $k$, and whether assessment errors are public or private. In Figures 4 and 5, we investigate the evolution of spatial patterns with different assessment rules. Finally, in Figure 6, we compare the results regarding the section with $b = 5$.

### 3.1. In the Absence of Indirect Reciprocity

For reference, we start by investigating how the regular ring lattice itself can affect the evolution of cooperation in the giving game. We consider only the Cooperator (ALLC) and Defector (ALLD). We conduct individual-based simulations in which in the initial setting of each node strategy on the graph is selected randomly between the ALLC and the ALLD, and other parameters are the same as in Figures 2 and 3. This investigation is independent of the quality of assessment errors because no strategy that depends on reputation assessment is assumed. We find that the spatial structure can only maintain cooperation at a very low rate (such as the mutation rate) under the typical parameter settings. No cluster of cooperation evolves for any degree of $k$. This indicates that the spatial structure itself would have no effect on the evolution of cooperation in the giving game.

### 3.2. Public Assessment

Figure 2 shows the results of indirect observation and public assessment errors. The results reveal that, among the four rules (shunning, image scoring, stern judging, and simple standing), stern judging and simple standing are most likely to promote cooperation and dominance by Discriminators. Similar to each other, and as the node degree of the network decreases, the threshold degree of $b$, across which stern judging or simple standing can lead to a full cooperation rate and Discriminator frequency, will increase. In stern judging and simple standing, sparse networks are more likely to facilitate the establishment of a prosocial state than dense networks.

In shunning, depending on the specific parameter settings, Discriminators can attain higher relative frequencies than in either stern judging or simple standing. However, this does not carry full cooperation (the cooperation rate is 0.6 at most in Figure 2A).

The cooperation rate in image scoring increases at a more gradual rate than in shunning because it reaches its maximum at a high benefit $b$ and middle node degree $k$ (about 0.8, as shown in Figure 2B); additionally, the Discriminator frequency only takes the intermediate value and thus does not move in correlation. It is observed that the Discriminator, ALLC, and ALLD can dynamically coexist within the network at non-large node degrees (Figure 4B). Particularly at low degrees of $k$, cyclical replacement occurs among the three strategies.

To better understand these phenomena in image scoring, we conduct extra simulations for both the Discriminator and the ALLD. These trials reveal that considering only the Discriminator and the ALLD leads the Discriminator to take over the entire population; however, there is an intermediate cooperation rate, which is independent of the node degree $k$, as in shunning. We note that both the Discriminator and the ALLC can achieve a higher cooperation rate in combination than in isolation on the regular ring lattice with a low node degree. Differently from the other three rules, image scoring cannot survive high node degrees in which population networks are highly dense. This is consistent with the results from the replicator dynamics of an infinite, well-mixed population [54].

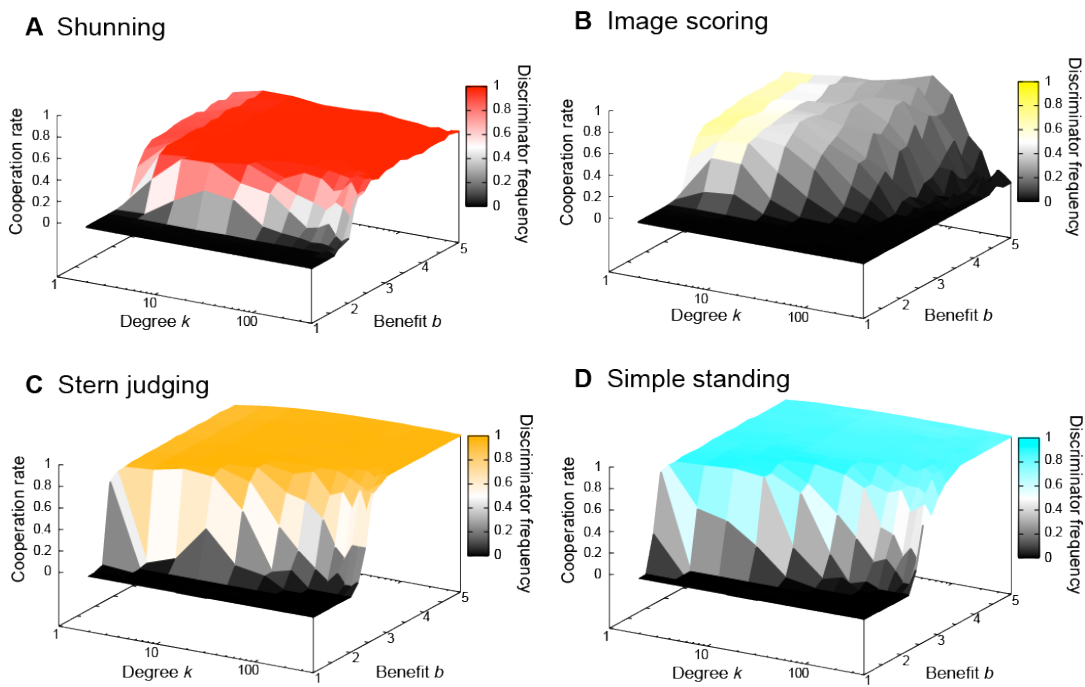

**Figure 2.** Cooperation rates and Discriminator frequencies in indirect observation and public assessment errors. (**A**) Shunning can take over the entire population yet achieve, at most, an intermediate cooperation rate; (**C**,**D**) both stern judging and simple standing can attain full cooperation in a state of almost full Discriminators for large benefits $b$; (**B**) image scoring can maintain a high cooperation rate with Discriminators whose frequencies are less than half. Parameters: $g$ = 500 generations, $h$ = 50 periods, error rates $p = q = 0.01$ (both for implementation and public assessment errors), selection intensity $s = 1$, mutation rate $m = 0.01$, node degrees = {2, 4, 8, 16, 32, 64, 100, 150, 200, 250, 300, 350, and 400}, $c = 1$, and $1 \leq b \leq 5$ (at interval 0.2). The cooperation rates and Discriminator frequencies depicted are the averages calculated over 20 independent runs of the agent-based simulation.

*3.3. Private Assessment*

Figure 3 shows the results of the direct observation and private assessment errors. The qualitative changes in errors have little effect on the resulting cooperation rates in simple standing and image scoring. However, this is not the case for shunning and stern judging. The horizontal surface of the maximal cooperation rate in shunning drastically drops to 0.2 in Figure 3B. The cooperation rate in stern judging suffers further catastrophic damage, decreasing all the way to zero and resulting in benefits *b* or high node degrees *k* in Figure 3C. In simulations with a longer period per generation, the cooperation rates in both shunning and stern judging further decline to zero in Figure 6B.

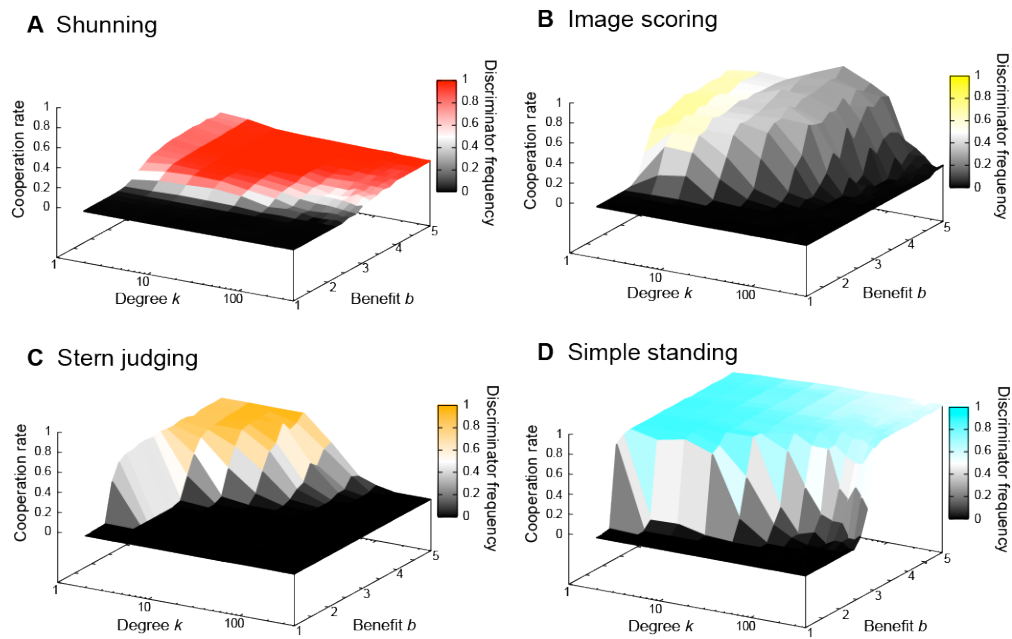

**Figure 3.** Cooperation rates and Discriminator frequencies in direct observation and private assessment errors. (**A**) Shunning can take over the entire population yet only achieve, at most, a low cooperation rate; (**C**) stern judging can achieve almost full cooperation, as can Discriminator frequency if, and only if, benefits *b* and node degrees *k* are sufficiently high and low, respectively; otherwise, both the cooperation rate and the Discriminator frequency reduce to zero; (**D**) simple standing can achieve full cooperation and 100% Discriminators for a broad range of parameters, as shown in Figure 2D; (**B**) image scoring can maintain a high cooperation rate in a mixed state with Cooperators and Discriminators (see also Figure 5B). Parameters: error rates $p = q = 0.01$ (both for implementation and private assessment errors) and other parameters are shown in Figure 2.

*3.4. Evolution of Spatial Patterns*

With no indirect reciprocity mechanism, Cooperators cannot survive by themselves in the presence of Defectors on the regular ring lattice (Section 3.1). Figures 4 and 5 show typical evolutionary patterns of spatial indirect reciprocity on the regular ring lattice, respectively, with public and assessment errors. For the lowest node degree 2, cycles of Discriminators (DISCs), Cooperators (ALLCs) (blue color), and Defectors (ALLDs) (black color) are observed throughout all four rules. With indirect observation and public assessment errors (Figure 4), shunning (SH) (red color) is most likely to dominate the population among the four rules; however, the cooperation level is not high. Stern judging (SJ) is likely to dominate as well. Notably, these results are not robust regarding changing within this kind of assessment error. With direct observation and private assessment errors (Figure 5), shunning leads to a low cooperation rate, and stern judging can result in both a low cooperation rate and Discriminator frequency. Compared with shunning and stern judging, the results from image scoring and simple standing are more robust for changes in both public and private cases. Image scoring (IS) is more likely to subsist than stern judging because, in forming clusters, it survives through a dynamic coexistence with ALLCs and ALLDs. The ALLD cluster is replaced with the Discriminator cluster; the Discriminator cluster will then be replaced with the ALLC cluster; the ALLC cluster will then be replaced with the ALLD cluster. For intermediate values of b and k, only image scoring can lead to a "rock-paper-scissors" type of dynamic. This yields a higher cooperation rate than that of the homogeneous states of Discriminators. When the node degree increases, the average frequency and cluster size of the Discriminator decreases. For larger benefits b and larger node degrees k, both the Discriminator and ALLC clusters can become finer, which can achieve the highest cooperation rate. Simple standing (ST) can maintain clusters of Discriminators for long periods, which can sometimes be replaced with ALLCs but not with ALLDs.

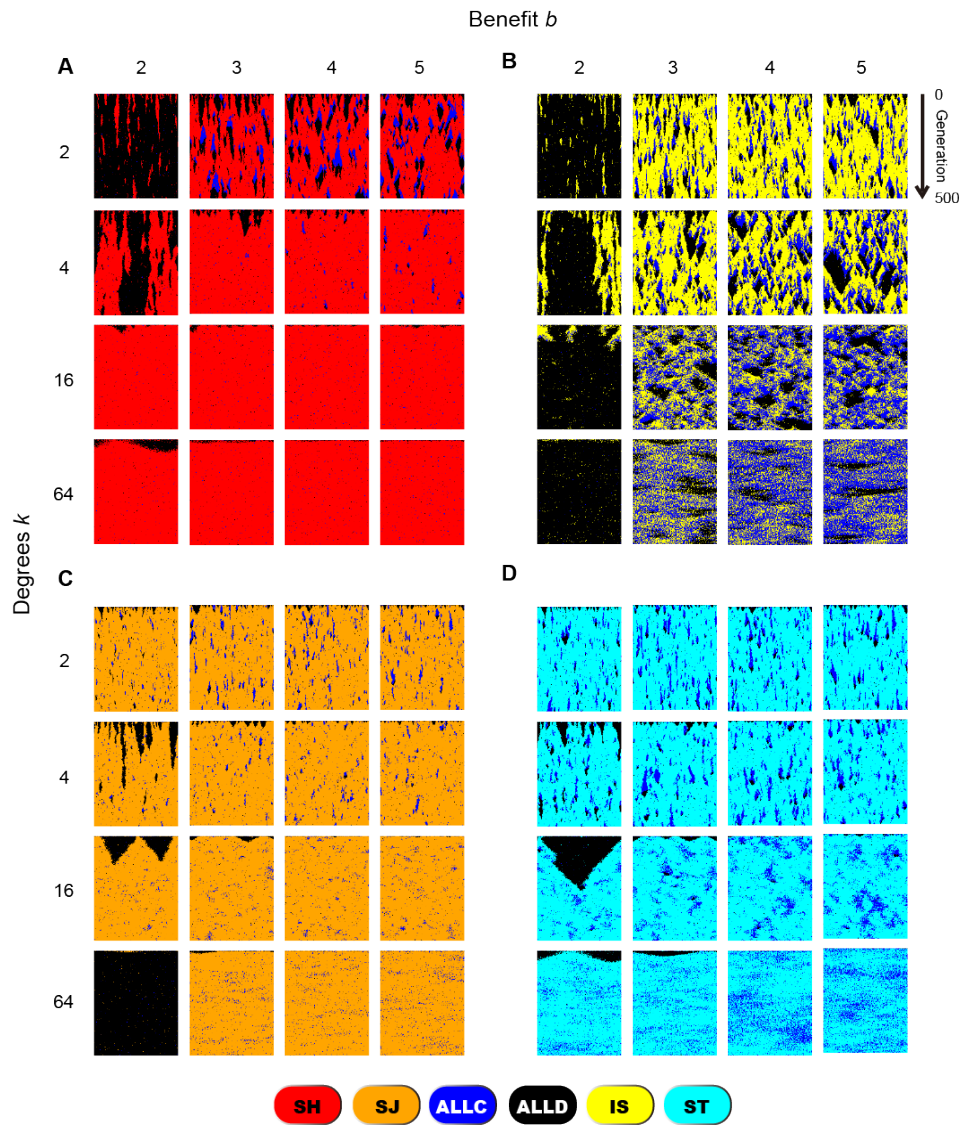

**Figure 4.** Evolution of spatial patterns of indirect reciprocity on regular ring lattices with indirect observation and public assessment errors. (**A**) Shunning (SH) (red color) and (**C**) stern judging (SJ) (orange color) are likely to take over the population. SJ can also achieve almost full cooperation; yet, in SH, the cooperation rate is low; (**B**) image scoring (IS) (yellow color) is most likely to lead to a three-strategy dynamical coexistence, in which case, the strategy is adopted by each node change frequently. As increases are seen in benefit $b$ and node degree $k$, the average cluster size of IS decreases, becoming replaced with an ALLC; (**D**) simple standing (ST) (cyan color) is most likely to maintain a state that is exclusively mixed with DISCs and ALLCs. Parameters: $g$ = 500 generations, $h$ = 50 periods, error rates $p = q = 0.01$ (both for implementation and public assessment errors), selection intensity $s = 1$, mutation rate $m = 0.01$, and $c = 1$.

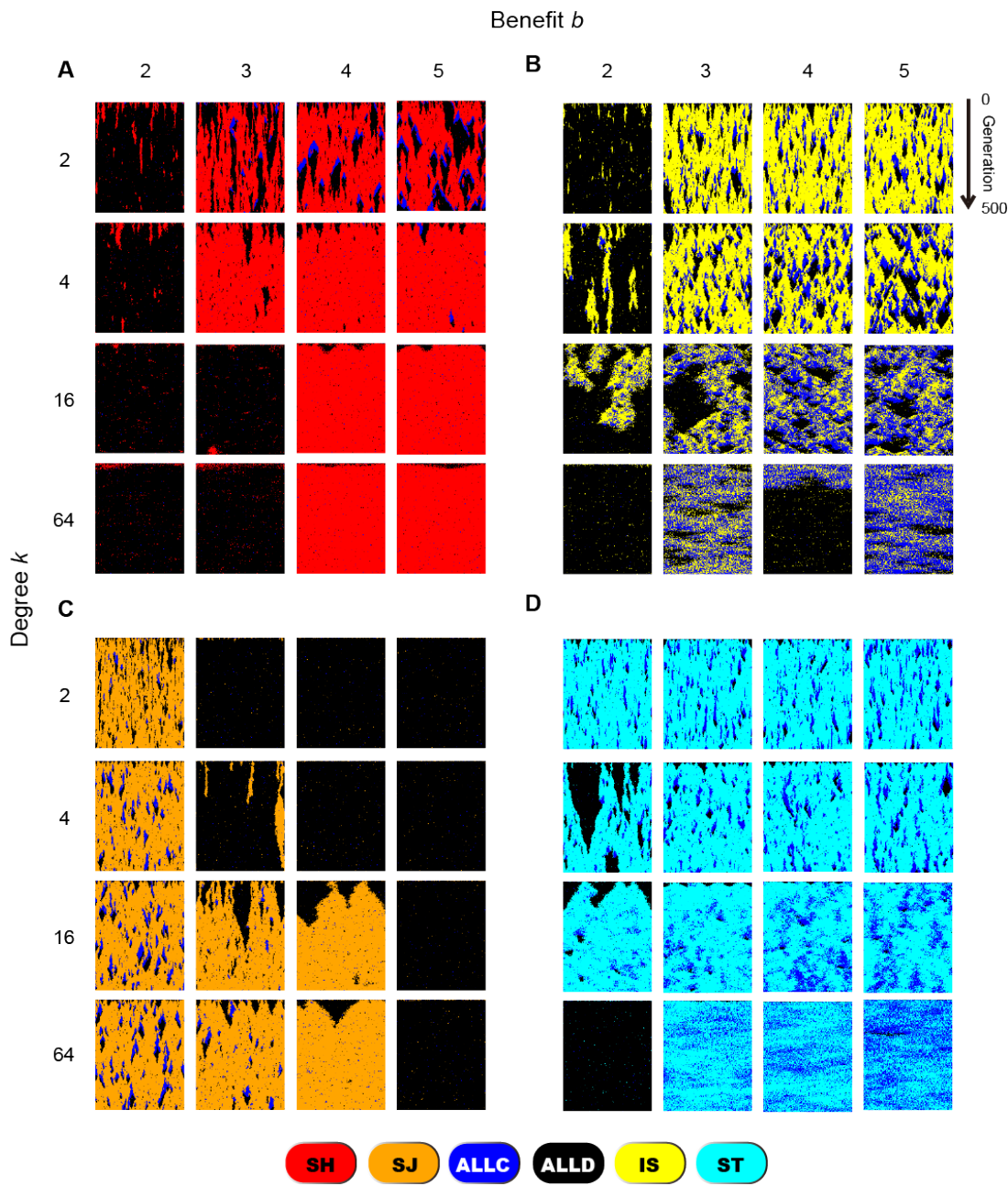

**Figure 5.** Evolution of spatial patterns of indirect reciprocity on regular ring lattices with direct observation and private assessment errors. (**A**) Shunning (SH) (red color) is likely to take over the population for the specific parameters; yet, the cooperation rate is low; (**C**) stern judging (SJ) (orange color) is less likely to thrive here than with the public assessment error, as shown in Figure 4C; (**B**) image scoring (IS) (yellow color) is most likely to lead to a three-strategy coexistence similar to what is shown in Figure 4B; (**D**) simple standing (ST) (cyan color) is most likely to maintain a state exclusively mixed of DISCs and ALLCs similar to what is shown in Figure 4D. The results from IS and ST remain qualitatively unchanged regarding the kind of assessment error. Parameters: $g$ = 500 generations, $h$ = 50 periods, error rates $p = q = 0.01$ (both for implementation and private assessment errors), selection intensity $s = 1$, mutation rate $m = 0.01$, and $c = 1$.

*3.5. Short Summary of Simulations Conducted for Long Periods*

The results in Figure 6 show how different node degrees and observation conditions affect the cooperation rate over the four rules and provide convenient information to compare shunning, image scoring, stern judging, and simple standing. In Figure 6, to clarify the effects of simulation periods, we consider 1000 periods per generation and compare them with 50 periods per generation in Figures 2–5. First, the regular ring lattice with small node degrees, which leads to a circle network, can achieve a substantial cooperation rate in almost all cases, except for shunning with direct observation and private assessment errors. Second, the regular ring lattice with large node degrees, which leads to a dense network, can only allow simple standing (in both cases with public and private assessment errors) and stern judging (only in the case with a public assessment error) to achieve a substantially high cooperation rate. The results from both shunning and stern judging are quite sensitive to the presence and quality of the assessment error. In private assessment errors, both shunning and stern judging lead to no cooperation (except for $k = 2$). In contrast to shunning, stern judging, and simple standing, the cooperation rate in image scoring takes its maximum value for an intermediate degree of $k$ and then declines as the node degree increases.

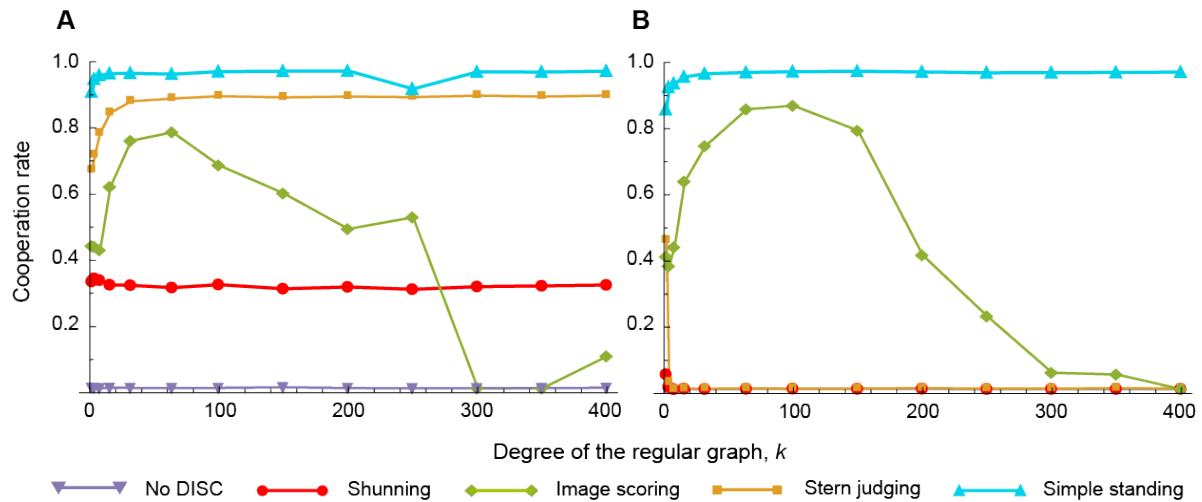

**Figure 6.** Effects of different node degrees on cooperation rates in simulations conducted for long periods. (**A**) With indirect observation and public assessment errors, shunning can only achieve the intermediate cooperation rate, and stern judging and simple standing can both consistently attain very high cooperation rates throughout almost all ranges of the node degree. Image scoring maximizes the cooperation rate at a node degree of around 64, which will decrease to zero as increases occur in the node degree. In the case with no DISCs, no cooperation evolves for any node degree. (**B**) With direct observation and private assessment errors, shunning achieves no cooperation for any node degree (except $k = 2$). Simple standing can maintain full cooperation, as shown in (**A**); however, in this case, stern judging cannot, which can lead to, at most, an intermediate cooperation rate only for $k = 2$ and an almost zero rate for any other node degree. Image scoring can lead populations to push the cooperation rate to its maximum for relatively small node degrees, which decreases, as shown in (**A**). Parameters: $g = 200$ generations, $h = 1000$ periods, implementation error rate $p = 0.01$, public and private error rates $q = 0.01$, selection intensity $s = 1$, mutation rate $m = 0.01$, degree = {2, 4, 8, 16, 32, 64, 100, 150, 200, 250, 300, 350, and 400}, $c = 1$, and $b = 5$. Cooperation rates depicted are the average calculated over 20 independent runs of the agent-based simulation.

## 4. Discussion

We explored reputation-based indirect reciprocity in regular ring lattices by specifically examining the four major assessment rules: shunning, image scoring, stern judging, and simple standing. Our results show that (i) simple standing, the most tolerant assessment rule among the four, can robustly maintain full cooperation over the cases of public and private errors; (ii) shunning can only achieve intermediate cooperation rates in the public case and no cooperation in the private case; (iii) stern judging can lead to full cooperation in the public case yet has one vulnerability, namely no cooperation when considering the private assessment error; and (iv) image scoring is sensitive to changes in the node degree. Our study is the first to use structured populations and observation conditions that clearly characterize the evolutionary dynamics of shunning, image scoring, stern judging, and simple standing. We particularly note that image scoring can achieve the highest cooperation rate through dynamic coexistence with ALLCs and ALLDs when considering a regular ring lattice with relatively small degrees. The spatial effects, as such, do not hold true in the other three assessment rules.

These assessment rules for indirect reciprocity were comparatively investigated and mainly based on well-mixed populations by means of the replication dynamics in infinite populations [57] or stochastic dynamics in finite populations [58]. Despite some differences in assumptions, our results with $k = N$ (complete graph), indirect observation, and public assessment errors are consistent with previous studies (particularly Figure 1 in Ref. [57] and the case of a population size of 300 in Figure 1 in Ref. [58]), as follows: for a relatively large benefit, (a) simple standing, stern judging, and shunning can take over the entire population, and image scoring cannot; and (b) simple standing and stern judging can lead to high cooperation rates, while shunning and image scoring can lead to low cooperation rates. Ref. [57] also discusses that, for shunning, it can take a significant amount of time for the cooperation rate to decline (to a low level), depending on the initial conditions and the error rates. This is confirmed in our individual-based simulations.

It is a challenge to fully understand the logic by which some of the assessment rules considered can promote cooperation in certain situations yet cannot in others. For instance, simple standing can maintain full cooperation throughout the public and private cases, and the cooperation rates in both shunning and stern judging decline to zero in the private case (Figure 6). Indeed, the analysis of indirect reciprocity with direct observation and private assessment errors has been a conundrum in the evolution of cooperation [59].

To develop further understanding of this issue, we focus on the dynamics of good image with direct observation and private assessment errors. According to our individual-based simulations, the probability that the Discriminator assesses another Discriminator, who is randomly selected, as good, would converge at around 100%, 0%, and 50% in (i) simple standing; (ii) shunning; and (iii) stern judging (after a sufficiently long period), respectively (see also Refs. [53,60]).

We first discuss (ii) shunning, noting that, without considering assessment errors, the shunning rule in implementation errors can only lead to almost all Discriminators having a bad image due to its strictness (Table 1). In indirect observation and public assessment errors, the updated assessment, whether right or wrong, is shared over the population. Thus, in shunning, even a small positive rate of assessment error can lead to maintaining at least some of the goodness rate in the public case. In the private case (direct observation and private assessment errors), the wrong assessment, such as erroneously assigning a good image, can occur only among a few Discriminators, which thus may be corrected in the subsequent periods. As a result, in the private case, shunning leads Discriminators to mutually assign a bad image.

Similarly, (i) simple standing with no assessment errors can lead almost all Discriminators to have a good image due to its tolerance (Table 1). In contrast to shunning, simple standing has a small positive rate of assessment errors, which have little effect on the maintenance of a high rate of good image for Discriminators in the public case as well as in the private case.

We then turn to (iii) stern judging, which—with indirect observation and public assessment errors—can maintain a goodness rate as high as that of simple standing. In stern judging, however, the indirect observation condition leads Discriminators to form two exclusive sub-groups that assess in-group members as good and out-group members as bad [60]. Our extra simulations confirm this and further reveal that the relative sizes of two exclusive sub-groups as such converge at around 50%. Considering that there is no bias among errors in the case of stern judging, the equalization of sub-group sizes seems plausible.

The dynamics of good image, as stated above, can lead to the evolutionary fate of cooperation in the private case. (i) In simple standing, the resident Discriminators, who mutually assign a good image and thus cooperate, are stable against the invasion of rare ALLDs because the ALLDs would be helped with a probability of 0% and thus worse off than the Discriminators with $b > c$. (ii) In contrast, in shunning, the resident Discriminators are not stable against the invasion of rare ALLDs because the Discriminators, who mutually assign a bad image and thus mutually defect, are not better off than the ALLDs. Similarly, (iii) in stern judging, the resident Discriminators are unstable against the invasion of rare ALLDs. We assume a sufficiently long period for each generation in numerical simulations such that the Discriminators assess each other as good with a probability of 50%. In the case of the rare ALLDs, an individual would be assessed as good with a probability of 50%. This leads to the conclusion that the rare ALLDs are better off than the resident Discriminators [60]. These analyses were conducted for the complete graph and could be applied to sufficiently large node degrees. We leave it to future work to analyze the dynamics of image and cooperation for small node degrees.

Studies of the evolution of reputation-based indirect reciprocity have mostly been based on a state in which the reputation distribution is already equal. In this study, we examined the two extreme cases of observation and dissemination: (i) indirect observation and public assessment and (ii) direct observation and private assessment. In (i), all the Discriminators equally shared the same assessment information provided by a unique observer. In (ii), all the Discriminators directly and independently observed and made assessments, with no consensus formation among one another. In reality, the intermediate case between these extreme cases, meaning probabilistic observation and local dissemination, would be more general. The extension to the intermediate case as such would also be useful for in-depth understanding of the characteristics of specific assessment rules, such as the sensitive results of stern judging, which depend on the presence of private assessment errors [55].

For the sake of simplicity, we have only considered a one-shot giving game, with no iteration of the game. However, our results could be applicable to the repeated prisoner's dilemma in regular networks, in which the probability to continue a round is sufficiently small. More generally speaking, our study indicates that the triple interplay of the three reciprocal mechanisms for the evolution of cooperation, including direct reciprocity, indirect reciprocity, and network reciprocity, can lead to fruitful outcomes. Our model has the potential to implement the interplay as such. To understand this, it is good to refer to the study of Observer Tit for Tat [40,61], a variant of normal Tit for Tat. Observer Tit for Tat differs from normal Tit for Tat only in the first move. In normal Tit for Tat, a player unconditionally cooperates in the first round and then reciprocates its partner by engaging in the same act last taken by this partner. Thus, normal Tit for Tat is worse off than unconditional defection (ALLD) due to the unilateral exploitation that occurs in the first round of ALLD. Observer Tit for Tat proposes defection in the first round if the partner's last action was found to be defection; otherwise, it proposes normal Tit for Tat action. Pollock and Dugatkin showed that Observer Tit for Tat can be better off than normal Tit for Tat when the fraction of ALLD is sufficiently large [61]. It is fair to say that Observer Tit for Tat corresponds to a combination of image scoring and normal Tit for Tat. Therefore, our results imply that an extensive combination of image scoring, normal Tit for Tat, and regular networks may be more likely to promote cooperation than random matching. Investigating the combination of direct and indirect reciprocity and spatial selection in this way would be an important step towards comprehensively understanding the different major mechanisms of the evolution of cooperation.

Another major reciprocal action against Defectors is peer punishment. Comparing cooperation withholding with peer punishment has gathered much attention from the viewpoint of indirect reciprocity in well-mixed populations [38,62,63]. Importantly, the mechanisms for punishment as well as higher-order moral assessment tend to be costly. Hence, this poses a second-order free rider problem—freeloading on others' efforts in making responsible punishment or assessments undermines the equilibrium of Punishers and Discriminators [44,64]. To address this problem, some solutions have been presented to date such as pool punishment with second-order punishment [65] and the punitive deposit system [66,67]. In indirect reciprocity, Sasaki and colleagues recently explored pre-assessment systems to detect second-order free riders [45]. In structured populations, in contrast to the case of well-mixed populations, costly peer punishment can be selected for without considering sanctions of second-order free riders [68]. Of interest would be a future study to compare withholding help against punishment in indirect reciprocity whilst considering structured populations and/or second-order free riders.

We have left out the two significant issues of scale-free network and deception. As is known, compared to regular networks, scale-free networks are more likely to help cooperation evolve [12,69,70]. Our preliminary results show that considering indirect reciprocity in scale-free networks can expand a range of parameters, in which a high level of cooperation is achieved. In addition, deception can lead to the destabilization of a prosocial state protected by indirect reciprocity [71–75]. Our results show that involuntary errors in making assessments can crucially affect the results of stern judging yet not those of simple standing. This might be applied to the consideration of the effects of deception in social networks.

**Acknowledgments:** We would like to thank the Austrian Science Fund (FWF): P27018-G11 (to T.S.), JSPS KAKENHI Grant Number 16H03120 (to I.O.), JSPS KAKENHI Grant Number 15KT0133 and 26330387 (to H.Y.). We would also like to acknowledge Satoshi Kurihara (University of Electro-Communications, Japan) for providing the computational resources.

**Author Contributions:** All authors conceived, designed, and analyzed the model, and revised the manuscript; H.Y. performed the numerical investigations; T.S. wrote the manuscript; and H.Y., I.O., and S.U. helped draft the manuscript.

**Conflicts of Interest:** The authors declare no conflict of interest. The founding sponsors had no role in the design of the study; in the collection, analyses, or interpretation of data; in the writing of the manuscript, and in the decision to publish the results.